\documentclass[
  ,final            % use final for the camera ready runs
  ]
  {aipproc}
\pdfoutput=1
\layoutstyle{6x9}

\begin{document}

\title{Do Nuclear Star Clusters and Supermassive Black Holes Follow the
Same Host-Galaxy Correlations?}

%\classification{<Replace this text with PACS numbers; choose from this list:
%                \texttt{http://www.aip.org/pacs/index.html}>}

\author{Peter Erwin}{
  address={Max-Planck-Institut f\"{u}r extraterrestrische Physik,
Giessenbachstrasse,
D-85748 Garching, Germany}
  ,altaddress={Universit\"{a}ts-Sternwarte M\"{u}nchen,
Scheinerstrasse 1,
D-81679 M\"{u}nchen, Germany}
}
\author{Dimitri Gadotti}{
  address={European Southern Observatory,
Alonso de Cordova 3107, Vitacura,
Casilla 19001, Santiago 19, Chile}
}

%\author{<author2>}{
%  address={<common address for author2 and author3>}
%}
%
%\author{<author3>}{
%  address={<common address for author2 and author3>}
%  ,altaddress={<author1 address>} % additional visiting address
%}

\begin{abstract}
Recent studies have suggested a strong correlation between the masses of
nuclear star clusters and their host galaxies, an extension of the known
correlations between supermassive black holes (SMBHs) and their host
galaxies. By focusing on disk galaxies with well-determined black hole
and nuclear cluster masses, we argue that there is \textit{not} a
universal ``central massive object'' correlation after all: careful
analysis shows that while SMBHs correlate better
with the stellar masses of the bulge components, nuclear star clusters
clearly correlate better with \textit{total} galaxy stellar mass.
\end{abstract}

\maketitle

%%%%%%%%%%%%%%%%%%%%%%%%%%%%%%%%%%%%%%%%%%%%
%% MAINMATTER
%%%%%%%%%%%%%%%%%%%%%%%%%%%%%%%%%%%%%%%%%%%%

\section{Introduction}

The past fifteen years have shown that essentially all massive galaxies
in the local universe harbor supermassive black holes (SMBHs, with
masses $M_{\bullet}$ of $\sim 10^{6}$--$10^{9} M_{\odot}$). The same period
has also shown that SMBH masses correlate quite strongly
with several global properties of the host galaxies, especially central
velocity dispersion \citep{ferrarese00,gebhardt00} and bulge luminosity
or mass \citep[e.g.,][]{marconi03,haring04}. The implication is that
the processes which drove galaxy growth and the processes which drove
black hole growth were intimately linked, perhaps even the \textit{same}
processes.

The same period has also seen the discovery that many galaxies,
particularly later-type spirals, host luminous nuclear star clusters
\citep[NSCs; e.g.,][]{carollo97,boker02}; see the review by \citet{boker08}.
Recently, several authors have suggested that nuclear clusters and
central SMBHs share the \textit{same} host-galaxy correlations: in
particular, that SMBHs and NSCs have the same correlation with bulge
luminosity/mass \citep{wehner06,ferrarese06,cote06,rossa06,balcells07}. 

There is, however, reason to be cautious about assuming a direct
SMBH-NSC analogy.  The samples of \citet{wehner06} and
\citet{ferrarese06} were almost entirely \textit{early-type} galaxies
-- ellipticals and dwarf ellipticals -- which are essentially
``pure bulge.''  But we know that SMBHs in \textit{spiral} galaxies
correlate better with the bulge, and \textit{not} with the total galaxy
mass or light \citep[e.g.,][]{kormendy-gebhardt01}.  And there have been
prior claims that nuclear star clusters correlate with \textit{total}
galaxy light \citep[e.g.,][]{carollo98}. So the question is: do nuclear
clusters \textit{in spiral galaxies} correlate with the bulge, or with
the whole galaxy?

\section{Data Sources}

For nuclear cluster masses, we prefer those which have been
\textit{dynamically} measured, both because this is the most direct
analog to SMBH masses and because it circumvents any possible problems
with multiple stellar populations, which can confound attempts to
estimate stellar masses from broadband colors.  The  masses are drawn
primarily from the sample of \citet{walcher05}, with additional data
from \citet{ho96}, \citet{boker99}, \citet{matthews99} and
\citet{gebhardt01}, and \citet{barth09}, along with preliminary
measurements from L. Colina (private comm.).  This yields a total of 14
galaxies, with Hubble types Sbc--Sm; the majority are Scd. We also
include 15 galaxies from \citet{rossa06}, where the masses are estimated
from spectroscopy; some of these are earlier spirals (Sa--Sb), but
most are Sc and later.

Although current studies suggest that the $M_{\bullet}$--$\sigma$ relation is
tighter, with less intrinsic scatter, than the $M_{\bullet}$--$M_{\rm bulge}$
relation \citep[e.g.,][]{gueltekin09,erwin-gadotti10}, velocity
dispersion is \textit{not} the ideal measure to use here, for the simple
reason that the central velocity dispersion in nuclear-cluster hosts is
almost always that of the cluster itself, and is used in determining the
cluster dynamical mass.  So we compare nuclear cluster masses with the
\textit{stellar mass} of host galaxy bulge, and with that of the entire
host galaxy.  It is from these comparisons, after all, that we can most
clearly see how SMBHs correlate with bulge mass and not with total
galaxy mass (Figure~\ref{fig:smbh}).

For comparison, we use the SMBH dataset and galaxy/bulge mass values
compiled by \citet{erwin-gadotti10}.  Bulge masses in that study
are determined by 2D image decomposition \citep[via the BUDDA software
package;][]{desouza04,gadotti08} which incorporates bulge and disk
components \textit{and} optional bars and central point sources
(accommodating both nuclear star clusters and AGN). Bulge/total ratios
from the decompositions are combined with $K$-band total magnitudes from
2MASS to get bulge $K$-band luminosities; we then use optical colors
from the literature to estimate stellar $M/L$ ratios via \citet{bell03}
and thus determine bulge (and also total) $M_{\star}$. Note that we
explicitly define ``bulge'' to be the ``photometric bulge'' -- that is,
the excess light/stellar mass which is not part of the disk, bar, or
nuclear star cluster.  We defer questions of how SMBH (or nuclear
cluster) mass relates to so-called ``pseudobulges'' versus ``classical
bulges'' \citep[e.g.,][]{hu08,nowak10} to a later analysis.

We are currently working on 2D decompositions for the nuclear cluster
galaxies, in cases where those are not already available; in the
meantime, we present results from 1-D bulge/disk decompositions,
always excluding the region dominated by the nuclear cluster itself (or,
alternately, fitting it as an additional component). We use published 2D
decompositions for some galaxies, such as those in \citet{laurikainen04}
and \citet{barth09}.  We do not expect the results to change
significantly when 2D decompositions are used for the whole sample.

\section{Comparing Black Holes and Star Clusters}

\begin{figure}
  \includegraphics[height=.26\textheight]{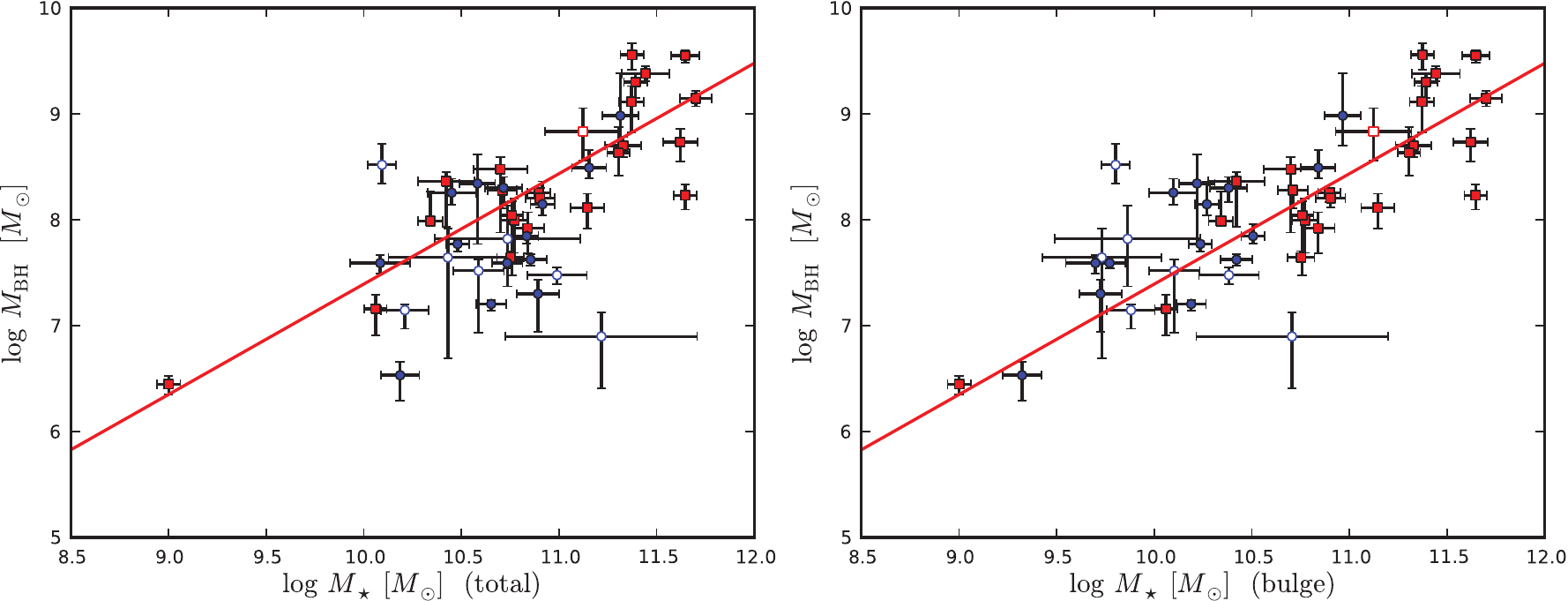}  

  \caption{Left: SMBH mass versus total galaxy stellar mass. Right: SMBH
  mass versus bulge stellar mass.  The diagonal line is the best fit to
  the elliptical galaxies (squares); open symbols are galaxies without precise
  distances. It is clear that the SMBH masses of S0 and spiral galaxies
  (circles) correlate better with the bulge stellar mass than with
  total galaxy mass. Based on \citet{erwin-gadotti10}.}\label{fig:smbh}

\end{figure}

Figure~\ref{fig:smbh} plots SMBH mass versus total and bulge stellar
mass. As is by now no surprise, the correlation of SMBH mass with
\textit{bulge} mass is much stronger than any correlation with total
galaxy mass (Spearman correlation coefficients $r_{S} = 0.71$ vs 0.29,
with the latter not statistically significant).

\begin{figure}
  \includegraphics[height=.26\textheight]{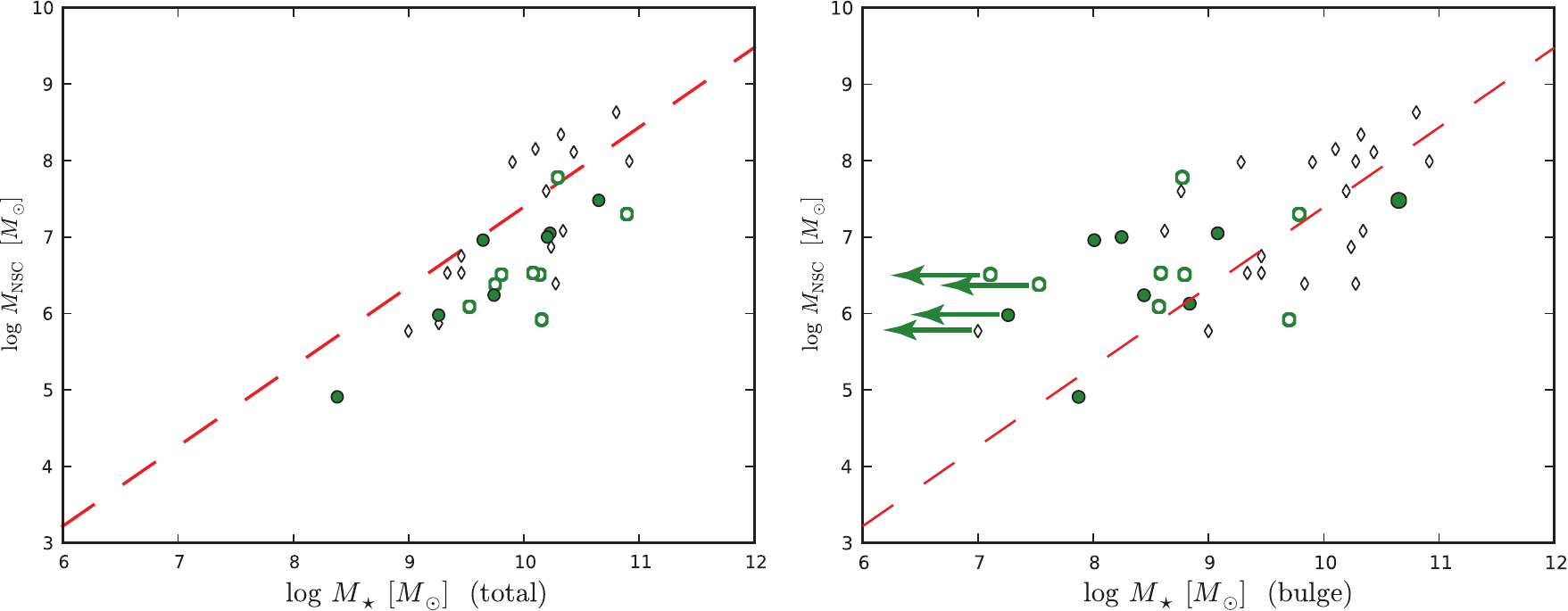}
  
  \caption{As for Figure~\ref{fig:smbh}, but now plotting
  \textit{nuclear star cluster} (NSC) mass versus total (left) and bulge
  (right) stellar mass; the diagonal dashed line is the SMBH-elliptical
  fit from Figure~\ref{fig:smbh}.  Diamonds are the spectroscopic-mass
  sample of \citet{rossa06}.  Arrows show nominal upper limits for four
  \textit{bulgeless} spirals (assuming $B/T \leq 0.01$). The situation
  is now the reverse of that for SMBHs: NSC masses clearly correlate
  better with \textit{total} galaxy mass than they do with bulge
  mass.}\label{fig:nsc-stellar-mass}
  
\end{figure}

The same plot for nuclear star clusters
(Figure~\ref{fig:nsc-stellar-mass}) shows the opposite: NSC mass clearly
correlates better with \textit{total} stellar mass than it does with
bulge mass: $r_{S} = 0.76$ versus 0.38, with the bulge-mass correlation
lacking any statistical significance.  The difference in correlation
coefficients actually \textit{understates} the contrast, because it
assumes that bulgeless spirals have nominal bulges ($B/T = 0.01$).  In
reality, the existence of nuclear star clusters in bulgeless spirals is
simply an unambiguous confirmation of the basic conclusion: nuclear star
cluster masses scale with the total stellar mass of their host galaxies,
not with the bulge mass, and are thus \textit{not} following the same
host-galaxy relation as SMBHs.

We have investigated whether other galaxy parameters might also
correlate with nuclear cluster mass, or with residuals from the $M_{\rm
NSC}$--$M_{\star}$ relation. In particular, we have compared nuclear
cluster mass with rotation velocity and total \textit{baryonic} mass. 
In both cases, correlations exist, but they are not as strong as the
correlation with total stellar mass.  So the latter appears to be the
defining relation between nuclear star clusters and their host galaxies.

%%%%%%%%%%%%%%%%%%%%%%%%%%%%%%%%%%%%%%%%%%%%%%%%
%% BACKMATTER
%%%%%%%%%%%%%%%%%%%%%%%%%%%%%%%%%%%%%%%%%%%%%%%%

\begin{theacknowledgments}
We acknowledge helpful conversations with Eva Noyola and Anil Seth.  This work 
was supported by Priority Programme 1177 of the Deutsche Forschungsgemeinschaft.
\end{theacknowledgments}

%%%%%%%%%%%%%%%%%%%%%%%%%%%%%%%%%%%%%%%%%%%%%%%%
%% The bibliography can be prepared using the BibTeX program or
%% manually.
%%
%% The code below assumes that BibTeX is used.  If the bibliography is
%% produced without BibTeX comment out the following lines and see the
%% aipguide.pdf for further information.
%%
%% For your convenience a manually coded example is appended
%% after the \end{document}
%%%%%%%%%%%%%%%%%%%%%%%%%%%%%%%%%%%%%%%%%%%%%%%%

\end{document}